# Advances in nano- and microscale NMR spectroscopy using diamond quantum sensors

Robin D. Allert[1], Karl D. Briegel[1], and Dominik B. Bucher[1,2]*

[1]Technical University of Munich, Department of Chemistry, Lichtenbergstr. 4, 85748 Garching b. München, Germany

[2]Munich Center for Quantum Science and Technology (MCQST), Schellingstr. 4, 80799 München, Germany

*Correspondence to dominik.bucher@tum.de

## Abstract

Quantum technologies have seen a rapid developmental surge over the last couple of years. Though often overshadowed by quantum computation, quantum sensors show tremendous potential for widespread applications in chemistry and biology. One system stands out in particular: the nitrogen-vacancy (NV) center in diamond, an atomic-sized sensor allowing the detection of nuclear magnetic resonance (NMR) signals at unprecedented length scales down to a single proton. In this article, we review the fundamentals of NV center-based quantum sensing and its distinct impact on nano- to microscale NMR spectroscopy. Furthermore, we highlight and discuss possible future applications of this novel technology ranging from energy research, material science, or single-cell biology, but also associated challenges of these rapidly developing NMR sensors.

## 1. Introduction

Since its inception, nuclear magnetic resonance (NMR) spectroscopy has inarguably become one of the workhorses of modern molecular structural analysis. Its significant advantages lie in its non-invasive structure elucidation while being quantitative and usable under standard conditions, making it ideal for synthetic chemistry, molecular biology, or material science[1]. However, the low intrinsic sensitivity of NMR spectroscopy typically restricts its application to macroscopic sample volumes, commonly hundreds of microliters, practically rendering it inapplicable to nano- and microscopic samples. Although, novel microfabricated coils have shown a promising avenue towards conventional inductively-detected NMR spectroscopy of sub-nanoliter volumes[2–5], they face challenges to further reduce the sample volume, e.g., difficulties of fabricating the microscopic pick-up coils and the proximity required to detect a signal from an even smaller number of spins.

In the last decade, a novel quantum sensor for magnetic fields has emerged, promising to overcome this sensitivity challenge. This sensor - the nitrogen-vacancy (NV) center - is one of the numerous color defects in diamond[6]. It attracted significant interest due to its spin state-dependent photoluminescence and straightforward spin manipulation enabling its use as a sensor for various physical quantities such as temperature, strain, and particularly interesting for chemists – oscillating magnetic fields, the decisive parameter in NMR spectroscopy[7–9]. Furthermore, due to its atomic size, the NV center can be brought into close proximity of the NMR sample, increasing signal strength and, accordingly, enabling NMR spectroscopy down to single cell volumes[10–12], single molecules[13–16] and even single spins[17–19].

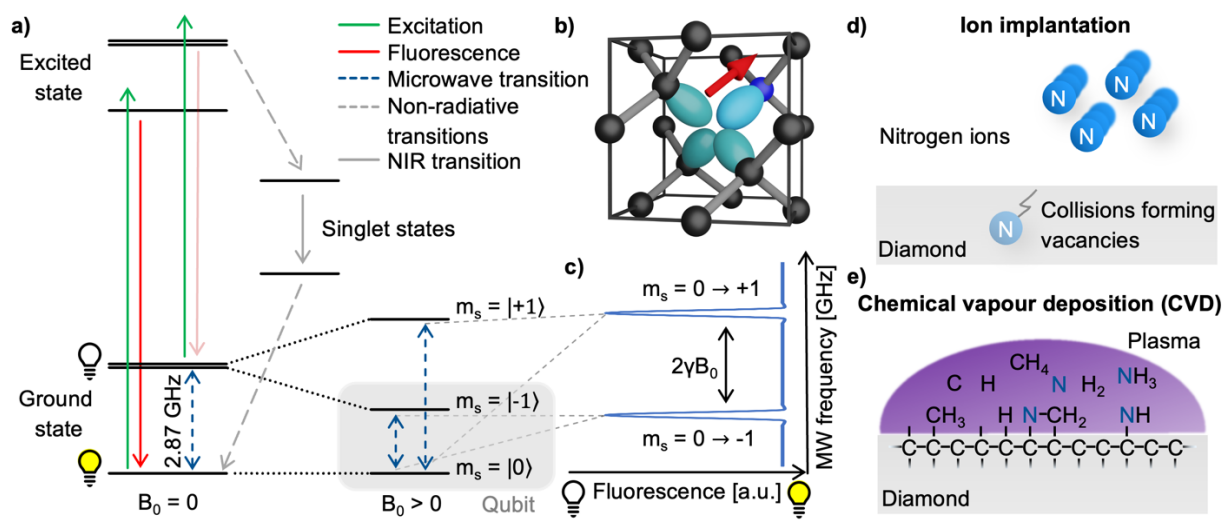

**Figure 1: Fundamental properties of the nitrogen-vacancy center.** a) Simplified energy level diagram of the nitrogen-vacancy (NV) center. Solid arrows denote radiative transitions (excitation in green and fluorescence in red), grey dashed arrows depict non-radiative transitions between the excited state, the singlet states, and the ground state. Only spin state conserving optical transitions are permitted, i.e., transitions between the same $m_s$ sublevels. Furthermore, the $m_s = |\pm 1\rangle$ excited states primarily decay through a non-radiative route via two singlet states leading to spin state-dependent fluorescence intensity and optical spin polarization of the $m_s = |0\rangle$ state. Additionally, the usually degenerated $m_s = |\pm 1\rangle$ states split in an external magnetic bias field (Zeeman splitting, $2\gamma B_0$ with $\gamma \approx 28$ GHz/T), which can be addressed individually with microwave transitions (blue dashed arrows). b) Crystal structure of the NV center; black spheres depict carbon atoms, the red arrow the orientation of the NV-center, and the blue sphere the corresponding nitrogen atom. c) Optically detected magnetic resonance (ODMR) experiment probing the $m_s = |0\rangle \rightarrow |-1\rangle$ and $m_s = |0\rangle \rightarrow |+1\rangle$ NV transition whose energetical splitting depends on the external magnetic field allowing for static and slowly varying magnetic field measurements. d) Nitrogen for NV center creation is commonly incorporated via ion implantation for shallow NV centers or e) directly during the diamond growth process (chemical vapor deposition) for micron thick NV-doped layers.

## 2. The nitrogen-vacancy center

### 2.1 The structure and formation of NV centers

The nitrogen-vacancy center (see Figure 1b) is a point defect in a diamond consisting of a nitrogen atom substituting a carbon atom (called P1 center) and an adjacent lattice vacancy[6,20]. The tetrahedral crystal structure of diamond allows four distinct NV orientations. Even though NV centers occur naturally, they are typically produced on purpose in highly purified synthetic diamonds to ensure high homogeneity and controllable properties for their application in quantum sensing[21,22]. Though different quantum sensing applications need specialized NV diamonds, the NV creation follows a general pathway of nitrogen incorporation, vacancy formation, and subsequent high-temperature annealing to form NV centers[23]. Generally, two methods of NV creation for NMR applications prevail: ion implantation and homoepitaxial diamond growth.

Nitrogen atoms are often incorporated into the diamond by ion implantation (see Figure 1d), where nitrogen ions are accelerated, penetrate the diamond's surface, and form P1 centers. Here, the ion's kinetic energy controls the penetration depth and, thus, the later NV depth (i.e. the distance from the surface)[21,24,25]. Furthermore, the total nitrogen ion fluence allows the subsequent NV density fine-tuning. During the implantation process, ions scatter in the lattice, colliding with carbon atoms - subsequently forming lattice vacancies which can diffuse efficiently inside a diamond at temperatures above 800 °C. In addition, the strain induced by the P1 centers causes the vacancies to diffuse towards the nitrogen[26] - accordingly forming NV centers upon combination[27]. However, the implantation process typically limits the NV-depth to tens of nanometers due to exceedingly high energies needed which causes unsustainable lattice damage.

Additionally, nitrogen-doped epitaxial-overgrowth of seed diamonds by chemical vapor deposition (CVD, see Figure 1e) allows the production of homogenous NV-layers with a wide range of nitrogen concentrations (several ppb to tens of ppm P1) and thicknesses (ranging from nanometers to hundreds of micrometers)[22,28]. Likewise, a small fraction of P1 centers (< 0.5%) already captures vacancies during the CVD growth due to lattice strain[29]. However, the NV concentration is significantly improved by creating additional vacancies, which can be achieved by high-energy irradiation with, e.g., electrons[30–32], ions[30,33], or lasers[34,35]. Like implanted diamonds, subsequent high-temperature annealing will cause the vacancies and P1 centers to form NV centers.

Commonly, CVD grown layers tend to exhibit more promising NV properties such as increased coherence times and NV yields than implanted samples[36,37]. Additionally, one significant advantage of CVD growth is the potential to form preferentially aligned NV ensembles along a single crystal axis[38–40]. However, CVD growth risks incorporating unwanted defects, e.g., NVH or crystal dislocations[22]. Furthermore, nanometer NV-layers by CVD growth are generally more challenging to realize and reproduce due to the complexity of the process[36,37]. On the contrary, implanting NV centers is a less complex process allowing for predictable nitrogen densities and depth distributions[41,42]. Further, single-ion implanters enable the deterministic placement of single NV centers[43–45].

The field of quantum material science has further developed various NV diamond types such as high-energy implantation with protons for micron layers[46], nanometer delta-doped layers[47], or the usage of high-temperature high-pressure (HPHT) diamonds[37]. For an in-depth discussion on color center creation, we would like to refer to an excellent review by *Smith et al.*[48].

## 2.2 Spin and photophysical properties of the NV center

The electronic structure of the NV center is governed by five electrons; where two electrons are provided by the lone pair of the nitrogen atom and three by carbon atoms adjacent to the lattice vacancy forming an electronic system around the vacancy comparable to molecular orbitals with long-lived spin states and well-defined optical transitions[49–51]. Several charge variants of the NV center exist; however, only the negatively charged NV⁻ is active for quantum sensing. Thus, our community uses the NV center synonymously with the NV⁻ charge state. The electronic ground state consists of the Zeeman states $m_s = |0\rangle$ and $|\pm1\rangle$ with a zero-field splitting of ~ 2.87 GHz; the latter are degenerate at zero magnetic field. After applying an external magnetic field $B_0$ along the NV axis, the $m_s = |\pm1\rangle$ states will split (Zeeman splitting, $2\gamma B_0$ with $\gamma \approx 28$ GHz/T). Crucially, the optical transitions between the ground state and the excited state are spin conserving, meaning that the optical transitions do not change the spin state $m_s$ of the involved electron (Figure 1a). Upon excitation of the NV center with light below ~ 640 nm (zero phonon line), the NV center has two relaxation pathways: either by fluorescence or a non-radiative pathway through intermediate singlet states which includes intersystem crossing. Notably, the $m_s = |\pm1\rangle$ excited states have a higher probability to relax through the non-radiative channel compared to the $m_s = |0\rangle$ state, leading to two crucial effects: first, the fluorescence rate of the $m_s = |\pm1\rangle$ states is reduced compared to the $m_s = |0\rangle$ states, and secondly, the optical excitation leads to spin polarization of the NV centers into the $m_s = |0\rangle$ state. These properties provide a convenient route for the spin state's preparation and read-out. Moreover, the transitions in the optical ground state between the $m_s = |0\rangle$ and the $|\pm1\rangle$ states can be addressed by microwave (MW) irradiation, permitting coherent manipulation of the NV centers spin state. In summary, the three vital NV center properties

1. **Efficient spin state preparation** to start in a well-defined quantum state,
2. **Coherent manipulation of the spin state** for controlled interaction with the environment, and
3. **Optical spin state read-out** to translate the spin state after the interaction with its environment to a measurable parameter,

provide a fully functional qubit, the basic unit of quantum technology, one of the few operational at room temperature[52]. While most branches of quantum technology try to isolate their fragile qubits as much as possible from the environment, quantum sensing hinges on the exceptional interaction of the qubit with its environment to perform measurements.

## 3. Quantum sensing

In general, quantum sensing refers to the process of employing a quantum system, often a quantum coherence, to measure a physical quantity - ideally with superior sensitivity, precision, or length scale compared to classical measurements[53].

### 3.1. Static magnetic field sensing using NV centers

The fundamental NV-based magnetic sensing scheme applies continuous excitation of the NV centers with laser light while sweeping the applied microwave frequency probing the transition of the $m_s = |0\rangle$ to $m_s = |\pm1\rangle$ state and, in turn, measures the strength of the applied magnetic field along the NV axis according to the Zeeman splitting ($2\gamma B_0$). At the start of the measurement, the NV center is in the bright $m_s = |0\rangle$ state due to optical polarization. However, if the MW frequency matches the transition $m_s = |0\rangle \rightarrow |\pm1\rangle$, the population of the $m_s = |0\rangle$ state decreases, thus the overall fluorescence intensity decreases (Figure 1c). Consequently, NV centers can translate magnetic fields into optical signals (optically detected magnetic resonance, ODMR), measured with a photodiode or camera, with widespread applications[8,54,55]. However, this scheme can only measure static (dc) or slowly varying (< 100 kHz) magnetic fields not applicable to NMR measurements where the detection of alternating (ac) magnetic fields in the MHz range is required.

## 3.2 Alternating magnetic field sensing using NV centers

To sense ac magnetic fields, we utilize well-established pulse sequences from classical NMR; however, rather than applying them to the sample nuclei, we apply the pulse sequence to our NV centers. In other words – NMR signals are detected via electron spin resonance (ESR) on the NV center. The general sensing scheme can be summarized as follows: First, the NV centers are optically polarized into the $m_s = |0\rangle$ state, then a MW pulse sequence is applied to coherently manipulate the NV spin state - usually to decouple the NV center from unwanted signals/noise and to enable the NV centers to selectively interact with the environment. To enable pulse sequences for NMR detection, a controllable two-level spin system (qubit) is prepared (see Figure 1a) by applying an external magnetic bias field $B_0$ along the NV centers axis to separate the $m_s = |\pm 1\rangle$ states. Which enables us to selectively address the $m_s = |0\rangle \rightarrow |-1\rangle$ or the $m_s = |0\rangle \rightarrow |+1\rangle$ transition with the respective states forming the poles of a Bloch sphere (from now on we will consider the $m_s = |0\rangle \rightarrow |-1\rangle$ transition, see Figure 2a). MW pulses control the position of the respective Bloch vector, and the quantum state can be read out optically via its spin-dependent fluorescence; therefore, translating the NMR signal into an optical signal.

Within this article, we would like to introduce briefly the field of quantum sensing pulse sequences and refer to several in-depth reviews on the topic of quantum sensing for further reading[53,55–57].

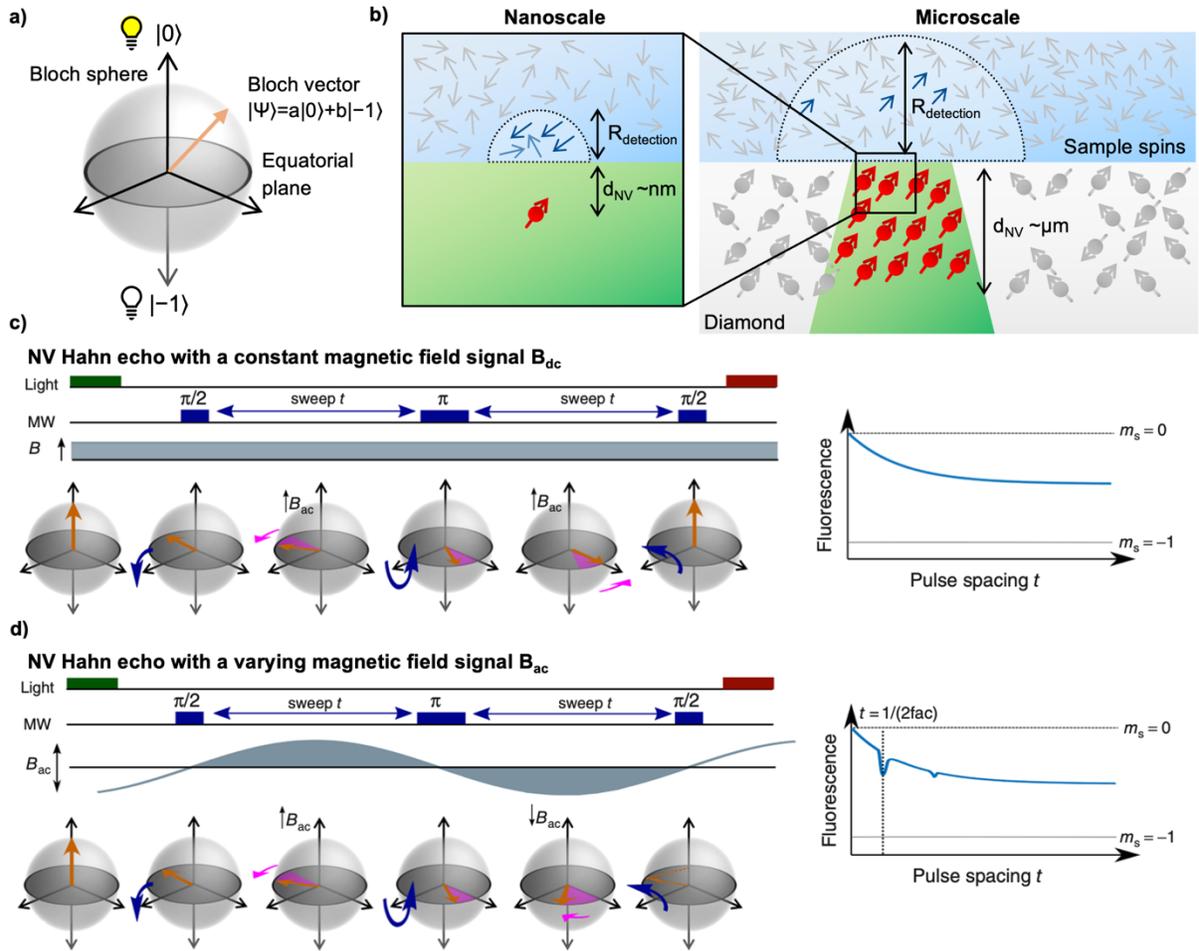

**Figure 2: Fundamentals of quantum sensing utilizing NV centers.** a) Bloch sphere picture of the NV centers two-level spin system where the bright $m_s = |0\rangle$ state is the north pole and the $m_s = |-1\rangle$ state the south pole. The orange arrow represents the Bloch vector, and the equatorial plane is depicted as a grey disk. b) Schematic of the detection radius of NV centers. Left: a typical nanoscale single NV experiment in which the NV depth and, thus, the resulting detection radius is in the range of a few nanometers (commonly below 10 nm), i.e., zeptoliters. Right: illustration of a typical microscale NV-NMR experiment utilizing an NV-doped diamond layer resulting in a detection radius of a few micrometers, i.e., picoliter detection volumes. The magnetic field detected by the NV-centers is originating from these volumes. c) NV Hahn echo experiment. The measurement sequence starts with a laser pulse to optically polarize the NV into the $m_s = |0\rangle$ state (Bloch vector is on the north pole), followed by a sequence of microwave (MW) pulses for spin control. The MW pulse sequence consists of π/2 - t - π - t- π/2 pulses, where the free precession time t between the pulses is swept. After the initial π/2 pulse, the Bloch vector will be in a superposition state on the equatorial plane. Here, the system will accumulate a phase (pink arrow) during the precession time t due to magnetic field variations (B), such as paramagnetic defects or other noise sources. The subsequent π pulse will mirror the Bloch vector within the Bloch sphere. For a constant noise source B, the phase accumulation during the second precession time will refocus the Bloch vector, i.e., cancel out the previously accumulated phase. Since magnetic noise is rarely constant, the Hahn echo will exhibit a decay setting the effective coherence time $T_2$. The conclusive π/2 pulse will translate the accumulated phase into a spin population, which is read out optically. d) NV Hahn echo with an external magnetic signal $B_{ac}$ (e.g., precessing nuclear spins in the detection volume). Suppose the time-varying signal matches the condition of $t = 1/(2f_{ac})$. In that case, the accumulated phase will not be cancelled but add up further because the sign of the magnetic signal has changed after the π pulse. Thus, the last π/2 does not map the Bloch vector back to $m_s = |0\rangle$, which results in a dip in fluorescence at $t = 1/(2f_{ac})$, effectively measuring the frequency of the external signal and, thus, allowing for NMR sensing with NV centers. Adapted from D. B. Bucher, *eMagRes* **2019**, *8*, 363–370 with permission from WILEY.

# Hahn echo experiment on the NV center

Most ac sensing pulse sequences are based on the well-known Hahn echo experiment in magnetic resonance spectroscopy. The Hahn echo pulse sequence starts with the NV center in the $m_s = |0\rangle$ state after the initial laser pulse (Figure 2c). Then, the first π/2 pulse (90° pulse) rotates the Bloch vector of the NV center into the equatorial plane of the Bloch sphere, which corresponds to a coherent superposition, e.g., $|\Psi\rangle = \frac{1}{\sqrt{2}}(|0\rangle + |-1\rangle)$. In the laboratory frame, the Bloch vector will rotate at its Larmor frequency around the z-axis. To simplify this picture, it is common in magnetic resonance spectroscopy to use a rotating frame revolving at the identical Larmor frequency, which leaves the Bloch vector static on the equatorial plane - we will do the same. Slight variations (i.e., magnetic noise B, see Figure 2c) in the magnetic field caused, for example, by $^{13}$C spins or other paramagnetic defects inside the diamond will induce small variations in the NV center's Larmor frequency, which forces the Bloch vector to precess on the equatorial plane. As a result, the Bloch vector will accumulate a total phase φ = γBt during the waiting time t (also called free precession time), yielding the state $|\Psi\rangle = \frac{1}{\sqrt{2}}(|0\rangle + e^{i\varphi}|-1\rangle)$. If a π pulse is applied, the Bloch vector will rotate by 180° around the applied axis, mirroring the Bloch vector on the Bloch sphere. If the magnetic noise is constant, the vector will proceed at its mirrored position in the same direction as before, effectively compensating the phase accumulation. However, magnetic noise is seldomly static, often varying in the time frame of our pulse sequences. Thus, the longer the free precession time t, the higher the probability that the accumulated phase by magnetic noise cannot be fully reversed. Finally, the second π/2 pulse will transfer this accumulated phase into an effective NV spin population which can be read out via its fluorescence. In a typical Hahn echo experiment, the precession time will be swept, while the fluorescence will be measured for every time point. For short times t, the spin population will solely be in the $m_s = |0\rangle$ state due to the neglectable phase accumulation. However, for ever-longer free precession times t, the NV center will accumulate phases due to random magnetic field fluctuations and transition to a mixed state somewhere between the $m_s = |0\rangle$ and the $m_s = |-1\rangle$ state. Therefore, the sweeping of the precession time will result in a decay curve of the measured NV center fluorescence, with the time constant, the coherence time $T_2$, of this decay being the practical sensing time set by the magnetic noise (spin bath) surrounding the NV center.

**AC sensing with dynamical decoupling sequences**

The properties of the NV Hahn echo experiment enable us to measure external ac magnetic fields caused, for example, by the Larmor precession of nuclei within the diamond or from a chemical sample outside of the diamond. The general method can be compared to the well-established electron spin echo envelope modulation (ESEEM) experiments known from classical ESR spectroscopy[58,59]. Typically, the influence on the Hahn echo will be minor if the free precession time t is much shorter or much longer than the period of the nuclear Larmor frequency (ac magnetic field). However, if the period $p_{NMR}$ of the ac frequency matches 2t, the phase accumulation will not cancel. Instead, the phase accumulation of the two free precession periods will add since the sign of the ac magnetic field has changed after the π pulse (Figure 2d). Thus, upon sweeping the precession time t, a decrease in the fluorescence at the time t corresponding to $p_{NMR}/2$ can be observed. This forms the basis for the majority of ac quantum sensing scheme and NV-NMR. The sensitivity and the spectral resolution of this scheme is limited by the coherence time of the NV center.

As already mentioned, the magnetic noise inside of diamond is mainly caused by other paramagnetic defects, ultimately limiting the coherence time $T_2$ and therefore the sensitivity. Thus, material engineering can advance ac sensitivity by employing high-purity synthetic diamonds (Type IIa) or even isotopically-engineered diamonds. However, another technique for enhancing the sensitivity is the application of multipulse sequences[60,61] based on the Carr-Purcell-Meiboom-Gill (CPMG) sequences known from conventional NMR spectroscopy. These sequences achieve more effective decoupling of the NV center from the ambient magnetic noise by applying additional π pulses on the NV center. Correspondingly, applying these π pulse trains will refocus the Bloch vector consecutively with every π pulse, and effectively prolonging the $T_2$ time, facilitating longer measurement times and, thus, higher sensitivities. Hence, the NV centers coherence time can reach up to milliseconds in room temperature[62] and hundreds of milliseconds in cryogenic temperatures[61]. After that, the experiment must be restarted, limiting the spectral resolution achievable to a few kHz. However, more sophisticated pulse sequences have been established, extending the practical measurement time, and reaching a theoretically arbitrary spectral resolution[10,63–66]. Interested readers are referred to excellent reviews[53,56,57]. Note that reliable information about the quantum state of the NV center can only be obtained by multiple readouts to gain sufficient statistics (quantum mechanics is intrinsically probabilistic, spin projection noise). Moreover, the readout fidelity is typically low due to limited number of detected photons per individual NV center (shot noise).

### 3.3 Detection volume of an NV-experiment

The key advantage of the NV center, compared to macroscopic induction coils, is its atomic size enabling close proximity to the sample nuclei. As a rule of thumb, a hemisphere on the diamond's surface where the radius equals the NV center's depth from the surface defines its NMR detection volume (Figure 2b)[67,68]. For nanoscale experiments the NV centers are close (a few nanometers) to the diamond surface. Whereas in microscale NV-NMR, micrometer thick NV ensembles are utilized. It is important to note, that an NV ensemble effectively acts as a single sensor with an average detection volume equal to the average depth. The detection volume itself determines in which polarization regime (i.e., nano- or microscale) NMR spectroscopy can be performed - see Box 1.

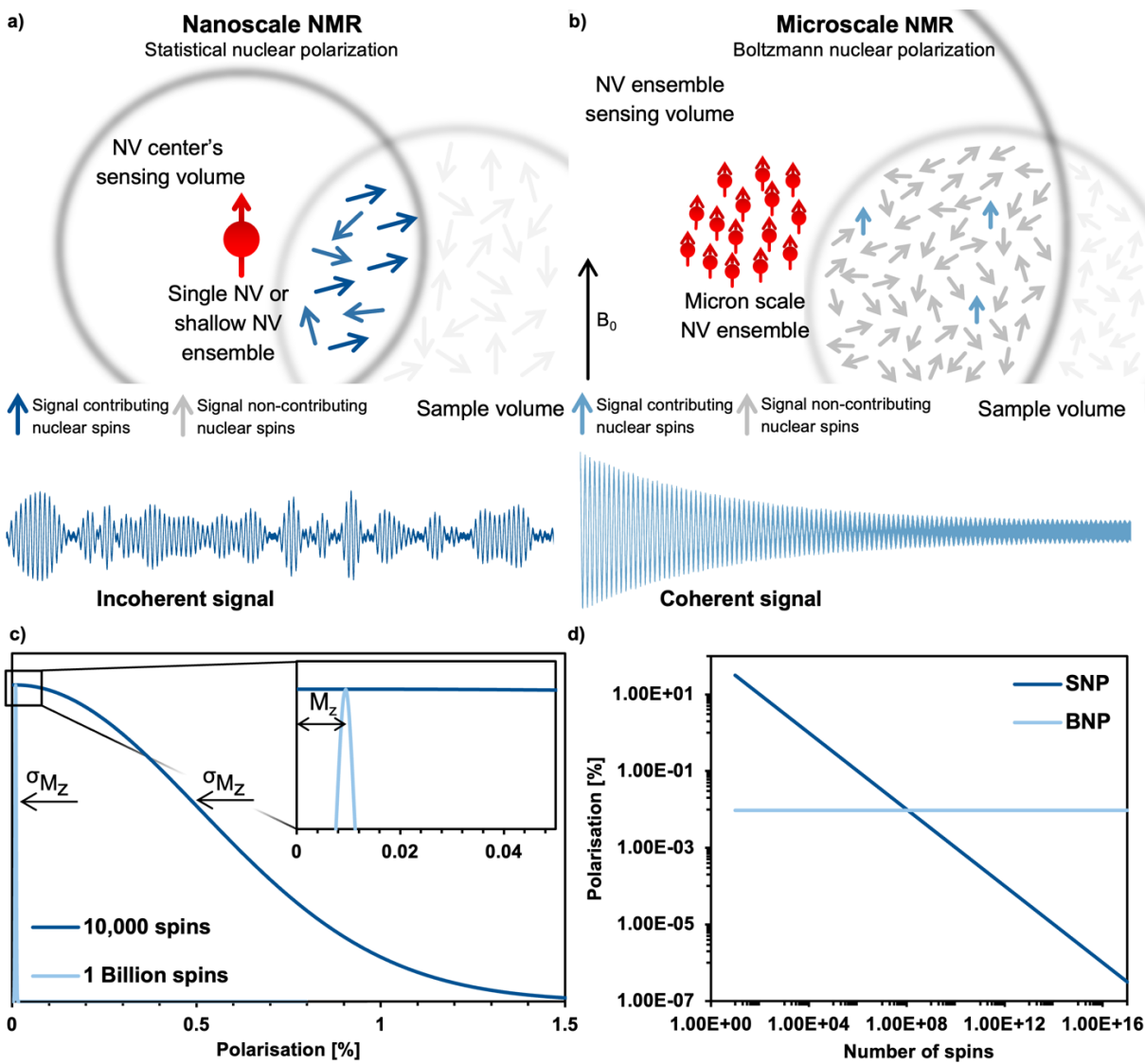

**Figure 3: Comparison of the two different regimes of NV-NMR:** a) Nano- (statistical nuclear polarization, SNP) and b) microscale (Boltzmann nuclear polarization, BNP) NMR. c) Schematic of a binominal distribution of the possible spin orientation (up and down) for ten thousand spins (SNP) and one billion spins (BNP) for protons at 28 T and 300 K. For small number of spins, random spin fluctuations ($\sigma_{M_z}$) result in a larger spin polarization than the mean $M_z$ (BNP) d) SNP and BNP as a function of number of proton spins at 28 T and 300 K. For small number of spins, SNP dominates.

**Box 1. Origin of the NMR signal in nano- and microscale NV-NMR experiments.** Microscale NV-NMR spectroscopy typically detects Boltzmann nuclear polarization – well-known from conventional NMR – where only minuscule fractions of the sample spins are polarized in the external magnetic field $B_0$ and contribute to the overall sample magnetization $M_z$ (see Figure 3c). This fraction is determined by the Boltzmann distribution and amounts to only ~ 100 parts-per-million for a 28 Tesla field and room temperature. Importantly, nanoscale NV-NMR experiments

detect only a small number of spins - here statistical fluctuations of spins lead to an incomplete cancellation of the magnetic moments; thus, resulting in an effective spin polarization of the chemical sample - called statistical nuclear polarization (SNP)[68,69]. *Müller et al.*[70] provide an intuitive picture of SNP based on a stochastic signal arising from a repeated quantized measurement of a nuclear spin ensemble: if a single spin (S = ½) is measured, its transverse angular momentum will result in $-\hbar/2$ for half the time and $+\hbar/2$ respectively for the other half. For two spins, the possible outcomes would be $\pm\hbar$ and 0. Therefore, for vast numbers of spins (N), the probability would approximate a Gaussian distribution with a standard deviation $\sigma_{Mz}$ of $+\hbar\sqrt{N}/2$. While the average of all measurements would result in 0, i.e., cancelation of the signal, the standard deviation $\sigma_{Mz}$ of the measured values is a nonzero result. Consequently, nanoscale NMR detects the variance ($\sigma_{Mz}^2$) of the magnetization rather than the sample magnetization $M_z$ (see Figure 3c inset).

Contrary to Boltzmann nuclear polarization (BNP), the statistical polarization scales with the square root of the inverse number of spins and can reach several percent polarization for small number of spins (see Figure 3c), while Boltzmann polarization benefits from higher magnetic fields and lower temperatures. Importantly, SNP detection does not require active excitation of the nuclear spins nor waiting for reaching equilibrium ($T_1$ time) for detection. However, it comes with some drawbacks, such as diffusion limited linewidths, discussed in the main text. Furthermore, we want to highlight the publication by *Schwartz et al.* examining the differences and applications of SNP and BNP in great detail[71].

**Table 1: Comparison of statistical and Boltzmann nuclear polarization.**

|  | **Nanoscale NMR** Statistical nuclear polarization (SNP) | **Microscale NMR** Boltzmann nuclear polarization (BNP) |
|---|---|---|
| Number of spins N | SNP $\propto \sqrt{1/N}$ | Independent |
| Magnetic field $B_0$ | Independent | BNP $\propto B_0$ |
| Temperature T | Independent | BNP $\propto 1/T$ |
| Achievable polarization | **Several %** | **~ 100 ppm at 28 T and 300 K** |
| Measured signal | No sample spin excitation necessary; Repetition time not limited by sample $T_1$ | Active sample spin excitation required (π/2 pulse); Repetition time limited by sample $T_1$ |

# 4. Nanoscale NMR

## 4.1. Fundamentals of nanoscale NMR

In a typical nanoscale NV-NMR experiment where the NV center is 2 - 10 nm below the diamond's surface, the NV, therefore, detects couple of hundreds to thousands of spins (1 nm$^3$ = ~60 protons in water). On these small length scales, the number of nuclear spins contributing to the NMR signal is not determined by the Boltzmann nuclear polarization (BNP) – but by random spin fluctuations (statistical nuclear polarization[68,69], SNP, see Box 1). This statistical nuclear polarization scales with one over the square root of the number of nuclear spins and reaches polarizations of up to a few percent resulting in relatively large magnetic signals at nearby NV centers. These signals can be easily detected with dynamic decoupling sequences, however not without a major caveat: nanoscale NMR does not yet reach spectral resolution sufficient for structure elucidation. Molecular diffusion is a substantial issue for liquid nanoscale NMR; water molecules, for example, diffuse through the detection volume of a single NV center in a few nanoseconds, effectively limiting the sensing time and, accordingly, broadens the linewidth (for water a few MHz to GHz)[72]. Thus, the diffusion of molecules determines the linewidth achievable in nanoscale NMR, currently restricting it to high viscosity/low diffusion constant samples, such as oils[67,73–75], polymers[67,76], or solids[73,77]. For the latter, anisotropic interactions such as dipolar couplings or chemical shift anisotropy induce spectral line broadening. This effect is well known from the field of solid-state NMR spectroscopy typically mitigated by magic angle spinning (MAS)[78]. However, MAS is technically challenging to implement for NV experiments.

## 4.2 Single NV experiments

The field of NV-NMR originated from solid-state physics studying single NV centers as nanoscale magnetometers offering a method to measure magnetic fields with sub-nanometer resolution. This field arose only a decade ago and since then has shown remarkable results detecting NMR signals from nanoscopic length scales. We want to highlight a few of them of particular interest for chemistry and biology.

In 2013, two landmark papers from *Staudacher et al.*[67] and *Mamin et al.*[76] successfully utilized a single NV center located roughly 7 nm under the diamond's surface to detect NMR signals from oils and poly(methyl methacrylate) (PMMA). Hereby, the depth of the NV corresponds to a sensing volume of only five cubic nanometers - ~13 zeptoliter - comparable to the size of a single mid-sized protein[67]. This work was expanded by *Loretz et al.*[75], demonstrating the measurement of $^1$H-NMR with an NV center ~2 nm below the surface equivalent to a measurement volume of eight yoctoliter ($8*10^{-24}$ L). Compared to state-of-the-art microcoil NMR (> 0.1 nL)[4], the required sample volume to generate a detectable signal was therefore reduced by twelve orders of magnitude.

These ultra-low volume detections were pushed to the limit by detecting the magnetic resonance of four silicon nuclei[17] in the strong coupling regime and later detecting a single proton[18]. In this regime, the interaction of the detected nuclei with the NV center is greater than the coupling to the surrounding other nuclei constituting an exceptional case for NMR in which the magnetic field of the individual spins is measured rather than the magnetization of the sample. Detection of a single proton has also been achieved by *Sushkov et al.*[19] by utilizing reporter spins (stable electron spins at the diamond surface) interacting with a proton and subsequently reading out these reporters by double electron-electron resonance (DEER) spectroscopy using a single NV center.

These proof-of-concept experiments were extended into chemically and biologically relevant samples in 2016. *Lovchinsky et al.*[13] demonstrated the detection of $^{13}$C-NMR and $^2$H-NMR signals form a single protein (see Figure 4a). The group utilized EDC/NHS crosslinking to covalently bind ubiquitin to the diamond surface and achieved co-localization of single proteins and single NV centers (see Figure 4b). Although highly impressive from a physics perspective, the practical applications are limited due to broad resonance lines in the dried solid-state they used for their measurement.

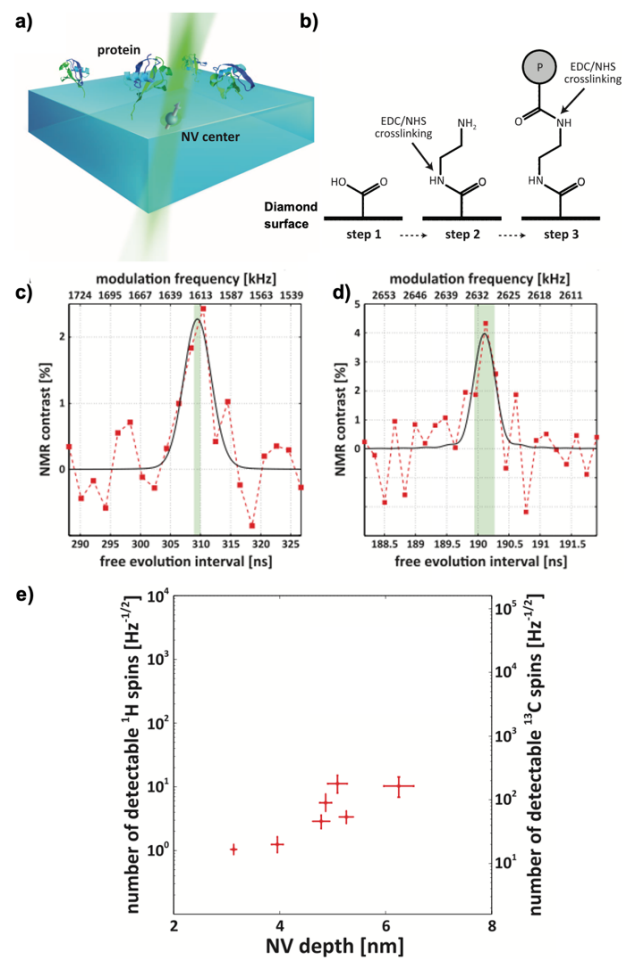

**Figure 4: NMR spectroscopy of single ubiquitin protein using NV centers.** a) Schematic depicting the experimental setup in which ubiquitin proteins were covalently tethered to the diamond's surface above a single NV center using EDC/NHS coupling. b) Schematic of the EDC/NHS crosslinking of the ubiquitin to the diamond surface (P stands for protein). NMR spectra of c) $^2$H and d) $^{13}$C originating from these single ubiquitin proteins. The red points represent measurement points and the respective Gaussian fit (solid black line). e) Measured depth-dependence of sensitivities for $^1$H and $^{13}$C NMR from single NV centers. All figures adapted from Lovchinsky et al., *Science* **2016,** *351*, 836-841. Reprinted with permission from AAAS.

Nevertheless, utilizing NMR to its full potential requires high spectral resolution - two essential parameters need to be resolved for structure elucidation: the *J*-coupling and the chemical shift. Neighboring nuclear spins cause *J*-coupling (typically in the range of 0.1-12 Hz for protons in small organic molecules independent of the magnetic bias field), mediated through chemical bonds, that offers information of the type, orientation, and number of close neighboring nuclei. Chemical shift is induced by the electron density of chemical groups close to the nuclei causing miniscule (ppm) magnetic field differences at the nuclear spin's location, modifying the local magnetic field and, thus, causing a relative shift of the resonance frequency in respect to the bias field $B_0$. Therefore, increasing the magnetic field $B_0$ is beneficial for chemical shift resolution. Consequently, NMR spectroscopy is preferably performed at high magnetic fields $B_0$; the implications for these high magnetic fields in NV-NMR experiments will be discussed in a later section. *Aslam et al.*[74] successfully demonstrated chemical shift resolution at the nanoscale utilizing a 3 Tesla superconducting magnet by using an advanced quantum sensing protocol which extended the read-out fidelity of the NV center by several orders of magnitude. This pulse sequence utilizes the intrinsic nitrogen nuclear spin of the NV center as a quantum memory, thus, improving the theoretically achievable linewidth resolution to 0.1 mHz or 0.01 ppb at 3 T for nanoscale NMR, in line with a modern conventional NMR spectrometer. However, as discussed earlier, the nanoscopic sample itself poses a fundamental limitation on the linewidth. *Aslam et al.* used highly viscous fluids as samples to minimize the effect of diffusion on the linewidth, which enabled them to measure chemical shift resolved $^1$H and $^{19}$F-spectra with a linewidth of 1.3 ppm and 1.4 ppm.

One promising route towards improving chemical information despite broad resonances in nanoscale NV-NMR is the detection of nuclear quadrupolar spins, i.e., spins with a spin quantum number greater than S = ½. The quadrupolar splitting is mediated by the interaction of the quadrupolar moment of the nucleus with the electric field gradient of the local electronic structure. These splittings are typically large (tens of kHz to several tens of MHz) and highly sensitive to the molecular structure. *Lovchinsky et al.*[79] detected quadrupolar resonances of $^{11}$B and $^{14}$N of atomically thin hexagonal boron nitride (hBN). The authors observed a dependence of the quadrupolar resonance with different layer thicknesses, paving the way to probe local structures in two-dimensional materials.

## 4.3 Shallow NV ensembles experiments

While single NV centers have a superior spatial resolution, there are certain drawbacks including tedious co-localization of the sample and a suitable NV center[13,16,80], intrinsically inhomogeneous properties[36,41,80] and instability of single, near-surface NV centers[81,82]. Furthermore, the detection of single photons make these experiments susceptible to background fluorescence. Consequently, the overall experimental complexity of single NV experiments[83] with long averaging times and background fluorescence results in low throughput for troubleshooting and/or detection of large parameter spaces. In our opinion, these drawbacks render single NV experiments so far difficult for applications in chemistry or biology. A straightforward method of effectively increasing the system's robustness, usability, and sensitivity is to employ an ensemble of NV centers close to the diamond's surface with the trade-off of losing the nanometer-scale resolution[84]. Instead, the diameter of the laser excitation - typically a few microns - and the depth of the NV centers define the measurement volume. This approach lends itself to simpler, robust, and highly sensitive setups, allowing easy NV-NMR integration into chemical or biological research.

*DeVience et al.*[73] demonstrated simultaneous NV-NMR of multiple nuclei ($^1$H, $^{19}$F, and $^{31}$P) with shallow NV ensembles and single NV centers. The ensemble approach circumvents the tedious search for a suitable NV center and allows for widefield magnetic resonance imaging (MRI) at the nanoscale imaged by a charge-coupled device (CCD) camera with a lateral resolution of ~500 nm and a field of view of 50 µm. Nanoscale imaging, where each pixel corresponds to an NMR spectrum, is one of the crucial benefits of the optical read-out of NV centers. The widefield NMR imaging approach was further expanded by *Ziem et al.*[85] by implementing optimal control sequences for the NV sensors, allow for more robust spin control even with inhomogeneous MW driving fields over the field of view. They applied their optimized pulses in magnetic resonance imaging of fluorine-containing thin films of 1.2 nm thickness resulting in an effective signal from ~120 nuclei per single NV center. Optimal control sequences are already applied in conventional NMR spectroscopy[86–88] as well as in NV-based sensing[89–91] demonstrating great potential to further increase sensitivity and robustness of the sensing schemes for NV-NMR spectroscopy.

So far, previous work detected the NMR signals directly on top of the diamond from either bulk samples, oils, or samples directly tethered to the diamond's surface. However, we envision the strength of NV-NMR in probing interfaces and surfaces beyond the diamond's intrinsic surface with applications especially for chemistry, which is one focus of our group.

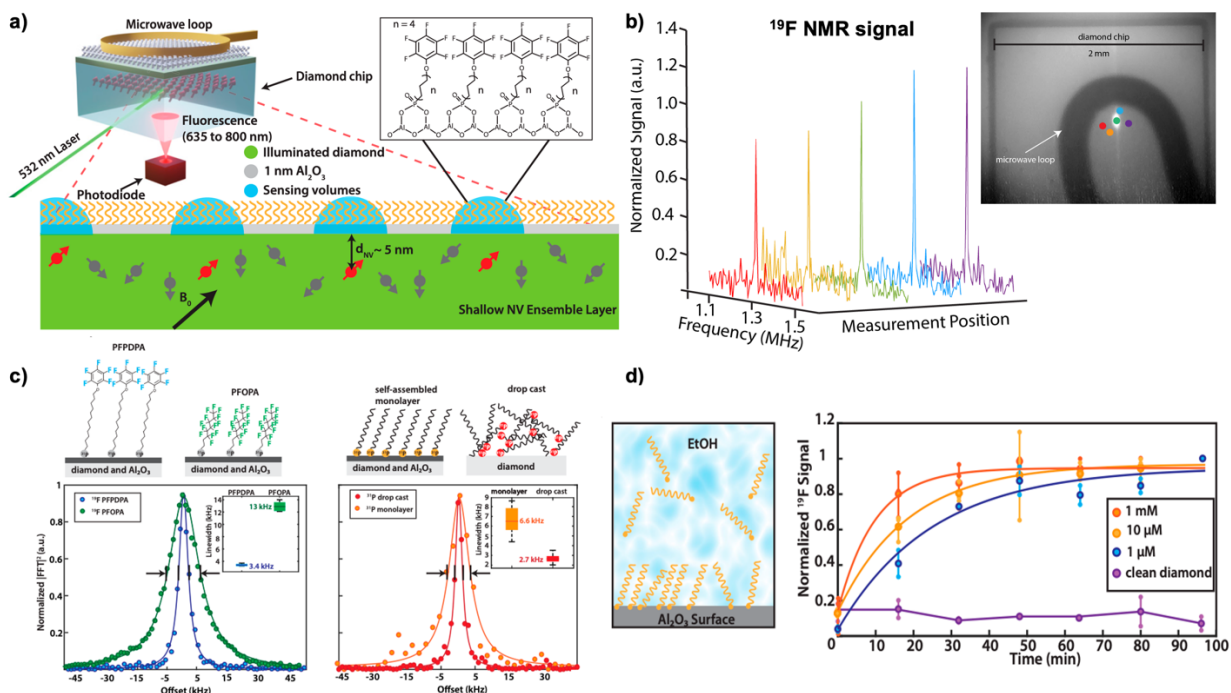

**Figure 5: Surface NMR using shallow NV ensembles for probing chemistry at interfaces.** a) Schematic of the surface NV-NMR experiments on a functionalized aluminum oxide ($Al_2O_3$) layer utilizing shallow NV ensembles, which measure the statistical polarization of fluorine and phosphorous from a self-assembled monolayer (SAM) of phosphonates. b) NV-NMR allows for spatial probing (imaging) of the SAM layer's homogeneity at several different measurement spots. c) Chemical information contained in the signal linewidth originating from the SAM can be detected, indicating local molecular dynamics and potentially the binding of the molecules to the aluminum oxide layer. d) Finally, NV-NMR allows for time-resolved in-situ studies of the monolayer formation under chemical relevant conditions observing the phosphonate-binding to the support structure, therefore, demonstrating different formation kinetics at various phosphonate concentrations. All figures adapted from K. S. Liu et al., *Proc Natl Acad Sci USA* **2022**, *119*, e2111607119 with permission from "Copyright (2022) National Academy of Science".

Interfacial and surface properties are vital for numerous material functions and applications in catalysis, energy conversion, or bioanalytics, especially under chemical-relevant conditions[77]. However, state-of-the-art analytic surface techniques require ultra-high vacuum, intense synchrotron radiation, or femtosecond lasers. Even then, chemical reaction kinetics and molecular dynamics are ultimately challenging to measure under ambient conditions. *Liu et al.*[77] recently demonstrated the application of NV-NMR in combination with atomic layer deposition (ALD) to study self-assembled monolayers (SAM) under ambient conditions (see Figure 5a). ALD is a technique that enables the surface deposition of thin films ranging from pure metals, common polymers (utilizing molecular layer deposition, MLD), or metal oxides[92]. In Liu's study, a one-nanometer thick amorphous aluminum oxide layer was deposited, allowing to probe the chemical modification of this layer with phosphonate chemistry during the formation of various self-assembled monolayers from organic compounds containing fluorine and phosphorus. The surface NMR technique detected both nuclei from the SAM-modified $Al_2O_3$ layer, providing chemical information with femtomole sensitivity with few tens of micron spatial resolution (see

Figure 5b), allowing for quantitative studies over the millimeter-sized diamond chip. Additional information on the SAM was obtained by comparing the linewidth of the fluorine signal (see Figure 5c). As discussed, before, in solid-state NMR spectroscopy, the linewidth is predominantly determined by dipolar coupling, which can be minimized by sample motion. Thus, fluorine located at the end of the SAM chain exhibits smaller linewidths due to their greater rotational freedom. Likewise, the line broadening effect can be used to study the binding effect of the phosphonate group to the $Al_2O_3$ support, where the $Al_2O_3$-bound moiety had significantly broader linewidths than drop-casted samples without a support layer, indicating chemical binding. In distinction to other analytical methods for surfaces, surface NMR allows time-resolved studies under chemical-relevant conditions enabling the observation of the real-time phosphonate-binding to the $Al_2O_3$ support at the solid-liquid interface (see Figure 5d). Therefore, NV-NMR enables direct detection of the reaction kinetics at the solid-liquid interface by measuring the NMR signal as a function of time, demonstrating faster SAM formation utilizing higher substrate concentrations.

## 4.4 Outlook

Even though single NV center studies offer superior spatial resolution and have reached the absolute detection limit of single spins, some drawbacks must be overcome to establish them as a general method such as sample and NV center co-localization. Over the past couple of years scanning NV experiments for static magnetic field sensing[93–95] have become a standard tool in solid-state physics and are commercially available. These scanning-probe experiments contain a scanning tip etched from high-purity diamond and include a single NV center. While typically used in nanoscale magnetic imaging, these microscopes could be adapted for NMR imaging - potentially solving the co-localization by bringing single NV centers to the sample allowing for nanoscale magnetic resonance imaging. In a similar fashion, the sample itself can be mounted on an AFM-like tip circumventing the colocalization issue[96,97]. These tip-based methods, however, face the potential challenge of increased distance between the sample and the respective NV center due to the inherent distance between the tip and the sample, reducing the signal strength. Another promising approach, NV-NMR with functionalized nanodiamonds containing NV centers has been successfully demonstrated[98] bringing the NMR detector into the sample itself and, thus, opening the route for NMR inside cells.[99]

By the same token, shallow NV ensembles for NMR spectroscopy have demonstrated their potential in surface sciences[77], due to the simple setup and robust measurements. Here, we envision a bright future for NV-NMR using NV centers as a novel analytical tool for material, catalysis, and energy research, ideally under in operando conditions.

A recent study has expanded the aluminum oxide platform for surface NMR demonstrated by *Liu et al.*[77], including additional binding moieties such as biotin or azides by multiple follow-up functionalization of the $Al_2O_3$ support[100]. This approach enables the efficient coupling of biomolecules such as proteins or DNA to the aluminum oxide deposited on the diamond's surface and expands the prospect of surface NV-NMR for biomolecules or biochemistry[101].

For all nanoscale NV-NMR experiments molecular structural information is presently limited by its diffusion-limited linewidth in liquid samples and dipolar coupling in solid-state samples to several kHz. Magic angle spinning known from classical NMR might solve the latter with significant engineering challenges. Furthermore, confinement of liquid samples for example in nanofluidics may restrict diffusion potentially enabling high spectral resolution[72,102]. Yet, recent results indicate that the diffusion can lead to sharp-peaked spectral lines which could enable improved resolution[103,104]. Another promising way to gain molecular information is the detection of quadrupolar nuclei (such as $^{11}B$, $^{14}N$, $^{2}H$, …), as discussed before[42,79].

## 5. Microscale NMR

### 5.1 Fundamentals of microscale NMR

While nanoscale NV-NMR greatly benefits from statistical polarization up to several percent, its diffusion-limited linewidth impedes its application in structural elucidation so far – one of the strong suits of NMR spectroscopy. Microscale NMR using dense NV layers with several micron thicknesses overcomes this issue by detecting the Boltzmann polarization. Effectively, microscale NV-NMR trades signal strength for spectral resolution. In this case, each equivalent nuclear spin of the sample behaves identical, and molecular diffusion does not limit the linewidth anymore. These experiments resemble conventional NMR experiments where a free nuclear precession (FNP) is induced by applying a π/2 pulse on the sample nuclei resulting in a coherent NMR signal.

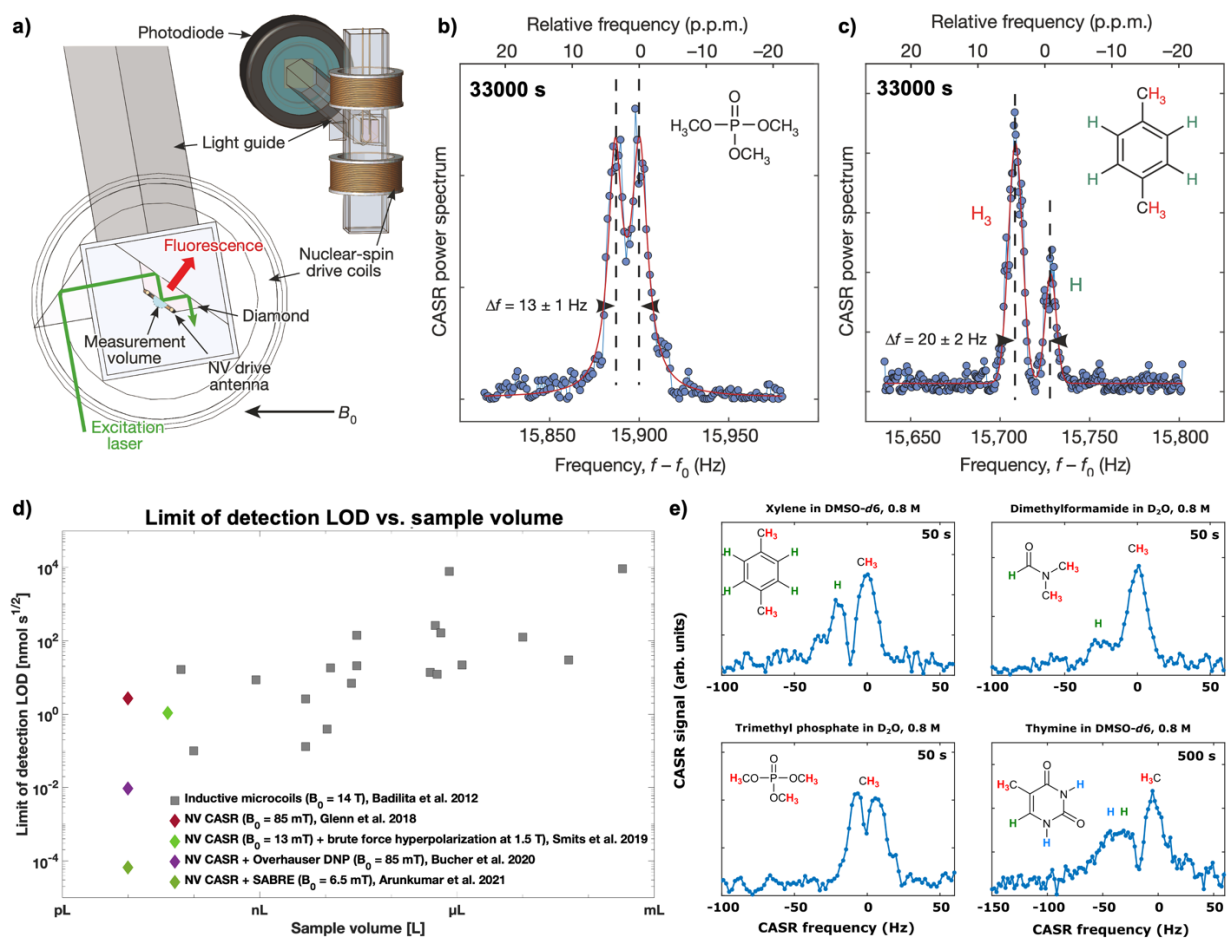

**Figure 6: Microscale NMR using high density NV layers with micrometer thicknesses for high spectral resolution.** a) Schematic of the experimental setup from Glenn et al.. The sample is placed inside a glass cuvette surrounding the diamond sensor, which is excited by a total internal refraction geometry to avoid sample illumination by intense laser light. Subsequently, a photodiode collects the fluorescence via a solid light guide. The sample's nuclear spins are resonantly driven by cylindrical coils surrounding the sample holder. Adapted from D. R. Glenn et al., *Nature* **2018**, *555*, 351–354 with permission from Springer Nature. b) CASR-detected NMR spectra of trimethyl phosphate measured at 88 mT. The dashed black lines depict the $^1$H-$^{31}$P *J*-coupling induced splitting of the NMR signal into a doublet. c) CASR-detected NMR spectra of xylene measured at 88 mT. The dashed black lines depict the chemical shift-induced splitting of the $^1$H-NMR signals. Both b) & c) were adapted from D. R. Glenn et al., *Nature* **2018**, *555*, 351–354 with permission from Springer Nature. d) Sensitivity comparison of microscale NMR detection methods, specifically the detection limit against the sample volume for inductive microcoils (gray, data adapted from V. Badilita et al., Soft Matter, **2012**, *8*, 10583.), NV CASR (red, data adapted from D. R. Glenn et al., *Nature* **2018**, *555*, 351–354.), NV CASR with brute force hyperpolarization (light green, data adapted from J. Smits et al. *Sci. Adv.* **2019**, *5*, eaaw7895.), NV CASR with Overhauser DNP (purple, data adapted from D. B. Bucher et al. *Phys. Rev. X* **2020**, *10*, 021053.), and NV CASR with PHIP SABRE (green, data adapted from N. Arunkumar et al. *PRX Quantum* **2021**, *2*, 010305.) e) $^1$H-NMR spectra of various small organic molecules measured with NV CASR at 88 mT using Overhauser DNP hyperpolarization demonstrating significant improvements in signal strength and, therefore, shorter measurement times compared to non-hyperpolarized samples. Adapted from D. B. Bucher et al., Phys. Rev. X 2020, 10, 021053 under CC BY 4.0.

## 5.2 High resolution microscale NV-NMR experiments

These signals are typically detected by the so-called coherently averaged synchronized read-out (CASR) sequence[10], meaning that dynamically decoupled sequences are synchronized to the detection signal (e.g. the FNP), allowing for arbitrary long measurement times and thus high spectral resolution. Recent works have demonstrated Hz-linewidths resolving *J*-couplings and chemical shift from chemical samples in picoliter volumes using CASR or adapted sequences.

Ultimately, the magnetic field uniformity and stability limited the spectral resolution of these experiments. In addition to the differences in linewidth and polarization, the sample geometry has a significant impact on the signal strength in microscale NV-NMR compared to the nanoscale. Similarly, the NMR signal strongly depends on the NV centers orientation concerning the diamond's surface – establishing a path for further improvements for the microscale NV-NMR setups. These effects have been investigated in great detail by *Bruckmaier et al.*[106].

The first principle study by *Glenn et al.*[10] has demonstrated $^1$H-NMR spectra resolving, for example, the distinct $^1$H-$^{31}$P *J*-coupling from trimethyl phosphate (see Figure 6b) and the chemical shift of protons in xylene (see Figure 6c) from a ten picoliter measurement volume – the volume of a single mammalian cell. Furthermore, even though long measurement times (~1 h) were required to achieve these spectra by solely measuring Boltzmann polarization, this study accomplished similar detection limits (LOD) with significantly lower sample volume than microscale inductive NMR (see Figure 6d). However, measuring Boltzmann polarization with NV-NMR is a double-edged sword: while the polarization and, thus, the detectable NMR signal and the chemical information benefits strongly from increased magnetic fields, NV quantum sensing becomes more and more challenging. The dynamic decoupling sequences utilized work most efficiently in the low MHz regime; fundamentally limited by the minimal achievable π pulse duration. Therefore, higher magnetic fields, i.e., higher NV resonance frequencies result in experimental complexity and raising the cost of the microwave equipment for the NV-spin control. For that reason, current microscale experiments have so far only been performed below 0.1 T. The impact of high magnetic fields on NV-NMR will be discussed in Section 5.4.

## 5.3 Hyperpolarization enhanced microscale NV-NMR experiments

To increase the concentration sensitivity, hyperpolarization methods known from conventional NMR or MRI have been successfully applied. First, in a brute-force hyperpolarization approach, the nuclear spin sample is pre-polarized at high magnetic fields, shuttled to the spectrometer then measured at low magnetic fields. *Smits et al.*[105] used this method to detect $^1$H-NMR and $^1$H-homonuclear and $^1$H-$^{19}$F-heteronuclear correlated spectroscopy (COSY) of 1,4-Difluorobenzene by pre-polarizing the sample at 1.5 T and subsequently measuring it at 13 mT, achieving one ppm linewidths.

Furthermore, microscale NV-NMR can be combined with auxiliary in-situ hyperpolarization techniques known from NMR such as parahydrogen-induced hyperpolarization (PHIP) or Overhauser dynamic nuclear polarization (DNP)[107]. Overhauser DNP specifically stands out for its ease of implementation in NV-NMR, providing signal enhancements of up to two orders of magnitude for $^1$H-NMR and up to three orders of magnitude for $^{13}$C nuclei[108]. This method requires the addition of stable radicals into the solution, typically a few mmol/L of TEMPO or its derivatives, which allows the transfer of Boltzmann polarization of the unpaired electron spin (~ three orders of magnitude higher than $^1$H polarization) to the nuclear spins of the sample by irradiating the sample with microwaves at the electron transition frequency. While integrating the Overhauser DNP in conventional NMR is usually not trivial because of the required microwave frequencies electronics for driving the electron transition[109], this is straightforward in NV-NMR because it already incorporates the necessary Gigahertz electronics for the NV spin control. Furthermore, Overhauser DNP has been shown to work best at low magnetic fields[110], facilitating NV-NMR. For example, *Bucher et al.*[11] showed femtomolar sensitivity (~5 mM concentrations) in a ~ ten picoliter volume. Correspondingly, they demonstrated enhancement factors between 80-230 times for a broad range of small molecules. Nevertheless, while straightforward to implement, Overhauser DNP has some adjoined disadvantages: the enhancement factor varies significantly between molecules[11] and even nuclei within the same molecule[110,111], complicating NMR studies. Moreover, the paramagnetic radicals cause line broadening[110], which may hamper the desired spectral resolution, and the Boltzmann polarization of the electron limits the maximum achievable enhancement.

Higher enhancements have been achieved via PHIP, where the nuclear singlet state of parahydrogen is used as a polarization source[112]. PHIP allows hyperpolarized molecules to be prepared by either hydrogenation reactions of unsaturated precursors (hydrogenative PHIP using

PASADENA[113] or ALTADENA[114] conditions) or by non-hydrogenative via an exchange process where the substrate and the parahydrogen exchange on the same metal center via signal amplification by reversible exchange (SABRE)[115,116]. *Arunkumar et al.*[12] combined the latter technique with microscale NV-NMR to achieve 0.5% proton spin polarization at 6.6 mT, improving five orders of magnitude over Boltzmann polarization. Accordingly, this method facilitated the measurement of small-molecule samples (e.g., $^{15}$N-labeled pyridine and nicotinamide) with concentrations as low as one millimolar from a ten picoliter volume, corresponding to 10 fmol, after 300 seconds of averaging time. While conventional NMR and MRI have shown a wide range of molecules polarizable with PHIP-based techniques, allowing for in-vivo pH sensing[117,118] or monitoring metabolites[119,120], PHIP is not universally applicable in chemistry or biology. However, the combination with the small volume detection of NV-NMR can open the field to single-cell analysis.

## 5.4 Outlook

Higher Boltzmann polarization is one of the primary motivations to reach for higher magnetic fields but not the only one; spectral resolution also dramatically benefits from higher magnetic fields splitting the NMR signals and, thus, allowing for better spectral resolution. So far, all NV-NMR experiments at the microscale have been performed below 0.1 T. However, magnetic fields below 0.35 T are an unfortunate regime for chemical analysis where the chemical shift and the *J*-coupling are at a similar magnitude (several Hz for protons). This "overlapping" of *J*-coupling and chemical shift induces strong couplings and second-order effects, which in turn cause the resonance lines to split at higher multiplicities than observed in typical high-field NMR. These effects can obscure the usually easy to interpret NMR spectra[10].

However, the NV center's resonance frequency at which the pulse sequence is performed and the respective NV center's π pulses scale with the magnetic bias field $B_0$. While the resonance frequency scales with ~ 28 GHz/T, the π pulses must be short enough to facilitate the high sample Larmor frequencies (tens of MHz) with dynamic decoupling sequences, i.e., significantly shorter than half the NMR frequency period. These form exceedingly difficult technical challenges requiring higher microwave frequencies and powers, which might cause among other dielectric sample heating and demand more expensive equipment. However, electron spin resonance (ESR) spectroscopy has solved some of these technical hurdles by establishing microwave resonators and amplifiers in the X band (8 GHz - 12 GHz) and $K_a$ band (26 - 40 GHz) corresponding to an NV-NMR range of approximately 0.2 T to 1.5 T. Likewise, magnets with sufficient homogeneity and stability in the sub-ppm range are widely available at these magnetic fields. Reaching these magnetic fields is one of the major goals of our community and would enable NMR spectra equivalent to modern commercial benchtop NMR devices, however, with significantly lower sample volumes.

NV-NMR is well suited for microscale measurements. Thus, further integration with microfluidics will provide an optimal match of the sample and measurement volumes. Chip-based microfluidics or Lab-on-a-Chip is an established platform for chemical[121] and biological applications . Its benefits hinge primarily on system miniaturization, such as well-understood mass and heat transport, as well as parallel processing, monitoring assays, and small amounts of reactants needed[123]. Current developments aim to reduce sample volumes further down to the picoliter regime[124]. While microfluidics can mimic most other established chemical or biological assay operations, e.g., mixing[125], incubation[126], or liquid-liquid extractions[127], its analytics is limited to fluorescence spectroscopy[128], absorption spectroscopy[129], or mass spectrometry[130,131]. However, the spectroscopic methods do not offer structural information and require either fluorescent samples or often complicated coupled reactions in the case of fluorescence spectroscopy. While mass spectrometry does offer structural information, it destroys the sample and is not generally applicable for every sample and solvent type. NV-NMR could overcome these limitations by offering non-invasive measurements of molecular structural information[105,132,133]. Furthermore, NV-NMR could be used to measure parameters such as molecular diffusion, flow rate, pH, or temperature inside the microfluidics, establishing a new analytical platform for Lab-on-a-chip applications.

Likewise, improvements in sensitivity and spectral resolution could enable NMR spectroscopy of small molecules and proteins in single cells with potential applications ranging from single-cell metabolomics[134,135] to NMR fingerprinting of tumors[135,136]. Furthermore, the optical read-out of NV-NMR allows for spatial resolution, which can be visualized as "scanning microscopy" by moving the laser excitation and thus the measurement volume. Combined with correlative optical microscopy, this would enable MRI measurements of tissues or biofilms.

## 6. Hyperpolarization using NV centers

So far, we have only discussed the NV center as a novel NMR sensor. However, the NV center itself has been proposed as a polarization source; optical pumping can rapidly (~μs) and efficiently (>80%) polarize the NV's electron spin. The resulting non-equilibrium polarization could be transferred to the chemical sample similar to classical DNP. However, unlike established DNP, there is no need for added paramagnetic radicals or cryogenic temperatures. Especially NV-doped nanodiamonds as polarization source for classical NMR demonstrate promising results[137] and have been realized using small form factor setups at room temperature[138], which could facilitate the practical translation of this technology. Therefore, this method of hyperpolarization could enable easy, low-cost, and efficient hyperpolarization for chemical analytics[132].

Furthermore, the combination of NV-based hyperpolarization with NV-detected NMR could enable applications requiring in situ hyperpolarization like single-cell NMR or NMR of nuclei with low gyromagnetic ratios. So far, only initial proof-of-principle studies have been conducted, demonstrating that polarization from single NV centers[139–141] and shallow NV ensembles can be transferred to spins outside of the diamond[142,143]. Nevertheless, there has not been a conclusive study demonstrating the subsequent detection of elevated polarization levels of sample spins outside of the diamond. Theoretical evaluations have proposed modest enhancements for microscale NV-NMR by one or two orders of magnitude with significant improvements by micro-structuring the diamond surface. In contrast, nanoscale NV-NMR is unlikely to benefit from this technology at the current stage due to the considerable statistical polarization[144].

Table 2: Outlook on the different regimes of NV-NMR and its potential application space.

| Regime | Field of study | Applications | Challenges | Solutions |
|---|---|---|---|---|
| **Nanoscale NMR** | Surfaces and interfaces | Battery research, electrochemistry catalysis, surface chemistry, bioanalytics, thin films, 2D materials, … | Limited spectral resolution | Nuclear quadrupolar resonance (NQR) spectroscopy[42,79] |
| | | | | Novel pulse sequences enabling detection at higher magnetic fields[145] |
| | | | | Nanofluidics sample confinement for diffusion restriction[72] |
| | | | | Magic angle spinning[78] |
| | | | | High magnetic fields (3 T) in combination advanced quantum sensing schemes[74] |
| | Single molecules | Structural biology (i.e., protein or DNA folding), catalysis, trace analysis, … | Sample co-localization | Nanofluidics[72], surface modifications[79,80,100] or scanning NV-NMR[96,97] |
| | | | NV charge state instability, leading to photoionization of the NV-centers | Co-doping with charge stabilizing electron donors[146,147] |
| | | | Limited throughput | Parallelized detection of single NV centers[148] |
| | | | Limited NV center homogeneity | Parallelized detection of single NV centers[148], material engineering[48] |
| **Microscale NMR** | Single cells | Single cell biology, metabolomics, tumor detection, directed evolution, … | Limited concentration sensitivity | Apply hyperpolarization methods such as Overhauser DNP[11], PHIP[12], dDNP, CIDNP, … |
| | | | | Improved NV read-out for sensitivity enhancement[149] |
| | | | Limited spectral resolution | Increase magnetic fields (> 0.35 T)[10] |
| | | | | Ultralow-field NMR in combination with hyperpolarization[150] |
| | | | | Novel pulse sequences enabling detection at higher magnetic fields[151] |
| | | | In vivo conditions (medium circulation, cell movement, …) | Microfluidic devices for cell cultures such as cell traps or nano titer arrays[152,153] |
| | | | Sample heating (Microwave & Laser heating) | Improved microwave resonator designs |
| | | | Single cell statistics | Magnetic resonance imaging[73,85] allowing the parallel detection of multiple single cells |
| | Lab-on-a-Chip | Nanomol chemistry, microdroplet screening, organ-on-a-chip, point-of-care testing, drug screening, … | Limited concentration sensitivity | Apply hyperpolarization methods such as Overhauser DNP[11], PHIP[12], dDNP, CIDNP, … |
| | | | | Improved NV read-out for sensitivity enhancement[149] |
| | | | Limited spectral resolution | Increase magnetic fields (> 0.35 T)[10] |
| | | | | Ultralow-field NMR in combination with hyperpolarization[150] |
| | | | | Novel pulse sequences enabling detection at higher magnetic fields[151] |
| | | | In vivo conditions (medium circulation, cell movement, …) | Microfluidic devices for cell cultures such as cell traps or nano titer arrays[152,153] |
| | | | Sample heating (Microwave & Laser heating) | Improved microwave resonator designs |

| | | | |
|---|---|---|---|
| **NV-Hyperpolarization** | Easy, low-cost, and efficient hyperpolarization for chemical analytics and diagnostics, … | Number of NV spins to sample spin ratio | Material engineering to increase the NV spin density[22,48] |
| | | | Nanostructuring of diamond to enhance surface area[144,154,155] |
| | | | Nanodiamonds containing NV centers[137,156,157] |

## Conclusion

The NV center in diamond has demonstrated its capabilities as a detector for nano- and microscale NMR spectroscopy. Researchers have realized impressive nanoscale NMR measurements of two-dimensional materials[79], single proteins[13], and even single protons[18,19]. However, low spectral resolution of nanoscale NV-NMR limits its applications so far. Since the technique is still in its infancy, future developments such as nuclear quadrupolar spectroscopy[79] or liquid sample[72] confinement will very likely overcome this drawback (see Table 2). Nevertheless, nanoscale NV-NMR employing shallow NV ensembles has shown its ability to measure NMR signals originating from surfaces and interfaces under chemical-relevant conditions expanding the toolbox of modern chemical analytics with widespread applications in catalysis, electrochemistry, and material research[77].

Microscale NV-NMR has made notable progress towards its application in single-cell studies or chemical analytics in microfluidics, demonstrating high spectral resolution and high spin sensitivity[10,66]. Furthermore, it can be readily interfaced with established NMR techniques such as hyperpolarization[11,12,105], 2D-NMR pulse sequences[105], or different nuclei[105], enabling structural and quantitative information at length scales of a single-cell (see Table 2). Additionally, microscale NV-NMR will benefit significantly from increased magnetic fields in the upcoming years, allowing for chemical shift resolution and sensitivity equivalent to commercial benchtop NMR systems – but with picoliter volumes.

We expect considerable progress in the field soon, establishing NV-NMR as a standard tool for chemistry and life sciences.

## Author Contributions

Conceptualization: R.D.A., D.B.B.; supervision: D.B.B; validation: R.D.A., K.D.B., D.B.B.; visualization: R.D.A., D.B.B.; writing – original draft: R.D.A., D.B.B.; writing – review & editing: R.D.A., D.B.B.

## Declaration of Interest

The authors declare that they have no known competing financial interests or personal relationships that could have appeared to influence the work reported in this feature article.

## Acknowledgments

This work was supported by the European Research Council (ERC) under the European Union's Horizon 2020 research and innovation program (grant agreement No 948049) and by the Deutsche Forschungsgemeinschaft (DFG, German Research Foundation) – 412351169 within the Emmy Noether program. The authors acknowledge support by the DFG under Germany's Excellence Strategy—EXC 2089/1—390776260 and the EXC-2111 390814868.